\documentclass[showpacs,prl,floatfix,twocolumn,amsmath,a4]{revtex4}
\usepackage{graphicx}
\begin{document} 
\title{Reply to  the comment by Ying Zhang and S. Das Sarma}
\author{S. De Palo} 
\affiliation{DEMOCRITOS INFM, Trieste, Italy} 
\affiliation{DFT, Univ.Trieste, Strada Costiera 11, 34014 Trieste,
  Italy}
\author{M. Botti}
\affiliation {CASPUR - Via dei Tizii, 6/b - 00185 Roma, Italy}
\author{S. Moroni}
\affiliation{DEMOCRITOS INFM, Trieste, Italy} 
\author{Gaetano Senatore} 
\affiliation{DEMOCRITOS INFM, Trieste, Italy} 
\affiliation{DFT, Univ.Trieste, Strada Costiera 11, 34014 Trieste,  Italy}
\begin{abstract} 
  We reply to the comment by Ying Zhang and S. Das Sarma on our PRL {\bf 94}, 226405 (2005).
\end{abstract}
\date{\today} 
\maketitle

\begin{figure} 
\includegraphics[height=85mm,angle=-90]{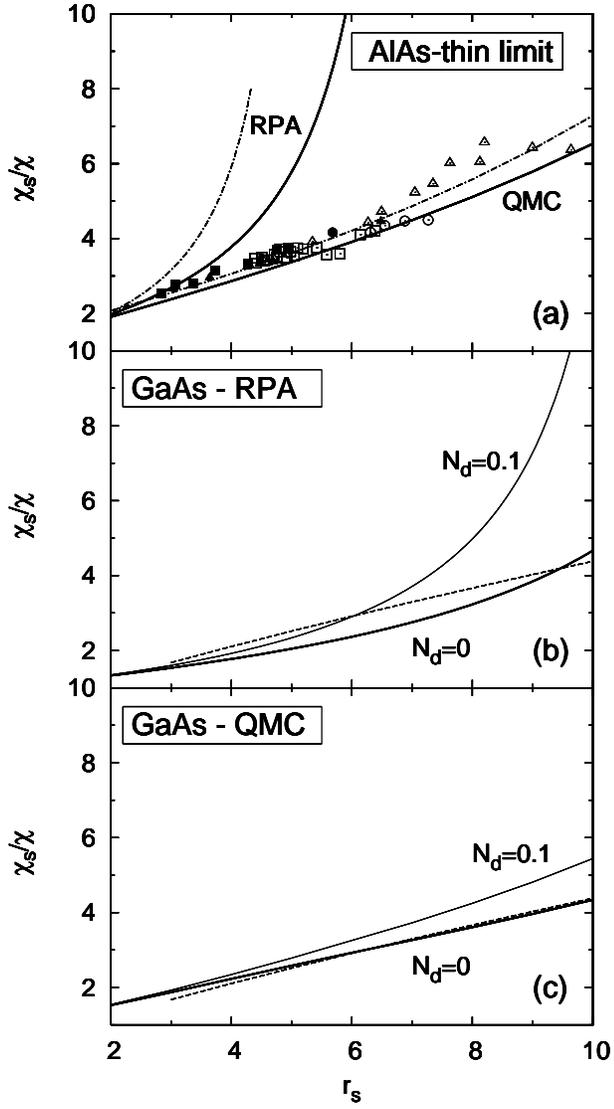}
\caption{Spin susceptibility of the 2DEG versus the coupling parameter
  $r_s=(\pi n)^{-1/2}/a_B$.  (a) theories (RPA\cite{prb}, upper two
  lines; QMC\cite{letter}, lower two lines) and
  experiment\cite{vakili} (symbols) in a thin EG (AlAs) ; (b) and (c)
  experiment\cite{zhu} (dashed curve) and theories (for two choices of
  the depletion charge density $N_d$ in units of $10^{11}cm^{-2}$) in
  a thick EG (GaAs).}
\label{chi}
\end{figure}

The agreement between our quantum Monte Carlo (QMC) based
predictions\cite{letter} of the spin susceptibility $\chi_s$ of the 2D
electron gas (EG) and the measurements in actual
devices\cite{zhu,vakili} is not accidental, as asserted in the
preceding Comment\cite{comment}. To assess the accuracy of theory, we
first focus on the thin limit of the 2DEG (ignored in \cite{comment}),
as realized in AlAs quantum wells\cite{vakili} with a transverse
thickness $t$ much smaller than the mean interelectronic
spacing\cite{letter}, $t\ll r_sa_B$. As is evident from Fig. \ref{chi}
(a), in this limit QMC is in very good agreement with experiment,
whereas the RPA theory of \cite{prb} largely overestimates the
experiment, being already off by a factor $\approx 2.5$ or larger at
$r_s=6$, hinting at an incipient divergence. For each of the
theoretical approaches displayed in Fig. \ref{chi} (a) the
experimental results should lay between the full and the dash-dotted
curves\cite{letter,prb}.

In the treatment of the thick limit, provided by the GaAs HIGFET of
\cite{zhu}, $t$ is determined by the depletion charge density $N_d$,
which in \cite{letter} was set to zero with the motivation given in
Ref. 18.  Here we can add that our calculation\cite{letter} of the low
density mobility in the HIGFET\cite{zhuthesis} yields an upper bound
to the unintentional dopant concentration in the GaAs $N_B=5 \times 10^{12}
cm^{-3}$, which considering the width $d=2\mu$m of the channel (the
GaAs region) turns into an upper bound for $N_d$, of $\approx
N_B\times d=0.01\times 10^{11} cm^{-2}$.  Such a low value of $N_d$ is in
our susceptibility calculations barely  distinguishable from the value
$N_d=0$, which in fact corresponds to our best fit of the measured
mobility\cite{zhuthesis} (whith a density of charged scatterers in the
AlGaAs of about $7\times 10^{13}cm^{-3}$).  Indeed, even with a value
$N_d=0.1\times 10^{11}$, which is completely ruled out by mobility
calculations, our prediction is still in fair agreement with
experiment, whereas the RPA prediction is already showing a tendency
to diverge not present in the experimental susceptibility, as is clear
from Fig. \ref{chi} (b)-(c).

Regarding other effects mentioned in \cite{comment}, we just note that (i) the
density dependence of band mass and g factor are found to be
negligible\cite{tan} for $r_s>1.5$; (ii) the Fermi temperature in GaAs
$T_F=(127/r_s^2)^oK$ is much larger then the temperature in the experiment
(T=30mK) at the densities considered; (iii) spin polarization effects have
been taken explicitly into account in \cite{letter}, through the calculation
of the {\it polarization field} susceptibility given by the dash-dotted curve
in the QMC result of Fig. \ref{chi} (a); (iv)finite transverse field $B_\bot$
effects are eliminated through extrapolation at $B=0$\cite{zhu,tan}.

Finally, Fig. \ref{chi} (a)-(c) clearly show that the agreement of our
QMC predictions with experiments is indeed very good and definitively
better than  that of RPA, which apparently and
substantially fails in the thin limit and performes somewhat better in
the thick limit ($N_d=0$) just because of the dominance of thickness.

\end{document}